\documentclass[aip,amsmath,amssymb,reprint]{revtex4-1} 
\usepackage{dcolumn}
\usepackage{bm}
\usepackage[colorlinks,hyperindex, urlcolor=blue, linkcolor=blue,citecolor=black, linkbordercolor={.7 .8 .8}]{hyperref}
\usepackage{graphicx}
\usepackage{tabularx}
\usepackage{amsfonts}
\usepackage{amsmath}
\usepackage{amssymb}
\usepackage{amsbsy}
\usepackage{setspace}
\usepackage{nicefrac}
\usepackage{xfrac}
\usepackage{bm}
\usepackage{float}

\newcommand\etal{{\it et\ al.}}

\begin{document}

\title{Extraordinary magnetoresistance in encapsulated monolayer graphene devices}

\author{Bowen Zhou}
		\affiliation{Department of Physics, Washington University in St. Louis, 1 Brookings Dr., St. Louis MO, 63130, USA.}
 
\author{K. Watanabe}
	\affiliation{National Institute of Materials Science, 1-2-1 Sengen, Tsukuba, Ibaraki 305-0044, Japan.}
    
\author{T. Taniguchi}
	\affiliation{National Institute of Materials Science, 1-2-1 Sengen, Tsukuba, Ibaraki 305-0044, Japan.}

\author{E. A. Henriksen}
  	\email{henriksen@wustl.edu.}
	\affiliation{Department of Physics, Washington University in St. Louis, 1 Brookings Dr., St. Louis MO, 63130, USA.}
   	\affiliation{Institute of Materials Science and Engineering, Washington University in St. Louis, 1 Brookings Dr., St. Louis MO 63130, USA}

\date{\today}

\begin{abstract}
We report a proof-of-concept study of extraordinary magnetoresistance (EMR) in devices of monolayer graphene encapsulated in hexagonal boron nitride, having metallic edge contacts and a central metal shunt. Extremely large EMR values, $MR=(R(B) - R_0) / R_0\sim 10^5$, are achieved in part because $R_0$ approaches or crosses zero as a function of the gate voltage, exceeding that achieved in high mobility bulk semiconductor devices. We highlight the sensitivity, $dR/dB$, which in two-terminal measurements is the highest yet reported for EMR devices, and in particular exceeds prior results in graphene-based devices by a factor of 20. An asymmetry in the zero-field transport is traced to the presence of $pn$-junctions at the graphene-metal shunt interface.
\end{abstract}

\maketitle

The extraordinary magnetoresistance (EMR) effect discovered by Solin and coworkers led to widespread interest in  this phenomenon for magnetic sensing applications \cite{solin_enhanced_2000,solin_nonmagnetic_2002,hewett_extraordinary_2012,sun_extraordinary_2013}. In the original work\ \cite{solin_enhanced_2000}, an extremely large magnetoresistance was found with the normalized  ratio $MR = \left(R(B) - R_0\right) / R_0$  as high as 16,000 \footnote{$MR$ values are often reported as a \% change. In this work we report the ratio itself.} at $B=5$ T, in devices comprised of a four-terminal InSb disk with a central metal shunt.  EMR is a geometric effect: at zero field, the shunt short circuits  current through the middle of the device. However in a perpendicular $B$ field the charge carriers are  deflected around the shunt as the current density $J$ becomes orthogonal to the interfacial electric field $E$ at the interface of the shunt and the semiconductor. Charge carriers are forced to travel through the more resistive material around the shunt, leading to a magnetoresistance enhanced over the zero field value by up to several orders of magnitude. Refinements to the original devices enable detectors suitable for fine-scale magnetic field sensors \cite{zhou_extraordinary_2001,solin_nonmagnetic_2002,sun_design_2012}. EMR  can be enhanced by tailoring the shape of the  shunt and also by asymmetries in the overall device configuration \cite{solin_extraordinary_2001,hewett_geometrically_2010,pugsley_extraordinary_2013}.

The advent of graphene device research in 2004 introduced a material with a readily  tunable, bipolar conductivity, and was soon followed by a first generation of graphene EMR devices \cite{novoselov_electric_2004,pisana_graphene_2010,pisana_tunable_2010,friedman_extraordinary_2011,lu_graphene_2011,el-ahmar_graphene-based_2017}. While an EMR effect was achieved, the $MR$ ratios remained below 1000,  well short of that in bulk semiconductors. Recently, significant improvement in graphene device quality has been achieved by encapsulating graphene in flakes of hexagonal boron nitride (hBN), an atomically-flat 6-eV-gap insulator \cite{kubota_deep_2007}. The hBN protects  graphene from extrinsic disorder, leading to very high quality  transport \cite{dean_boron_2010,wang_one-dimensional_2013}. With  device mobility   linked to greater EMR \cite{solin_enhanced_2000,hewett_extraordinary_2012}, here we investigate EMR devices fabricated from encapsulated graphene. For comparison, the InSb devices used by Solin \etal\ typically have mobilities of $\mu \approx 70,000$ cm$^2$/Vs\ \cite{solin_enhanced_2000};  graphene-on-oxide devices are  in the 1,000 to 10,000 cm$^2$/Vs range but the mobilities of hBN-encapsulated devices can be 2 or 3 orders of magnitude larger\ \cite{pisana_graphene_2010,pisana_tunable_2010,friedman_extraordinary_2011,lu_graphene_2011,el-ahmar_graphene-based_2017,dean_boron_2010,wang_one-dimensional_2013}.

\begin{figure*}[t]
\includegraphics[width=0.9\textwidth]{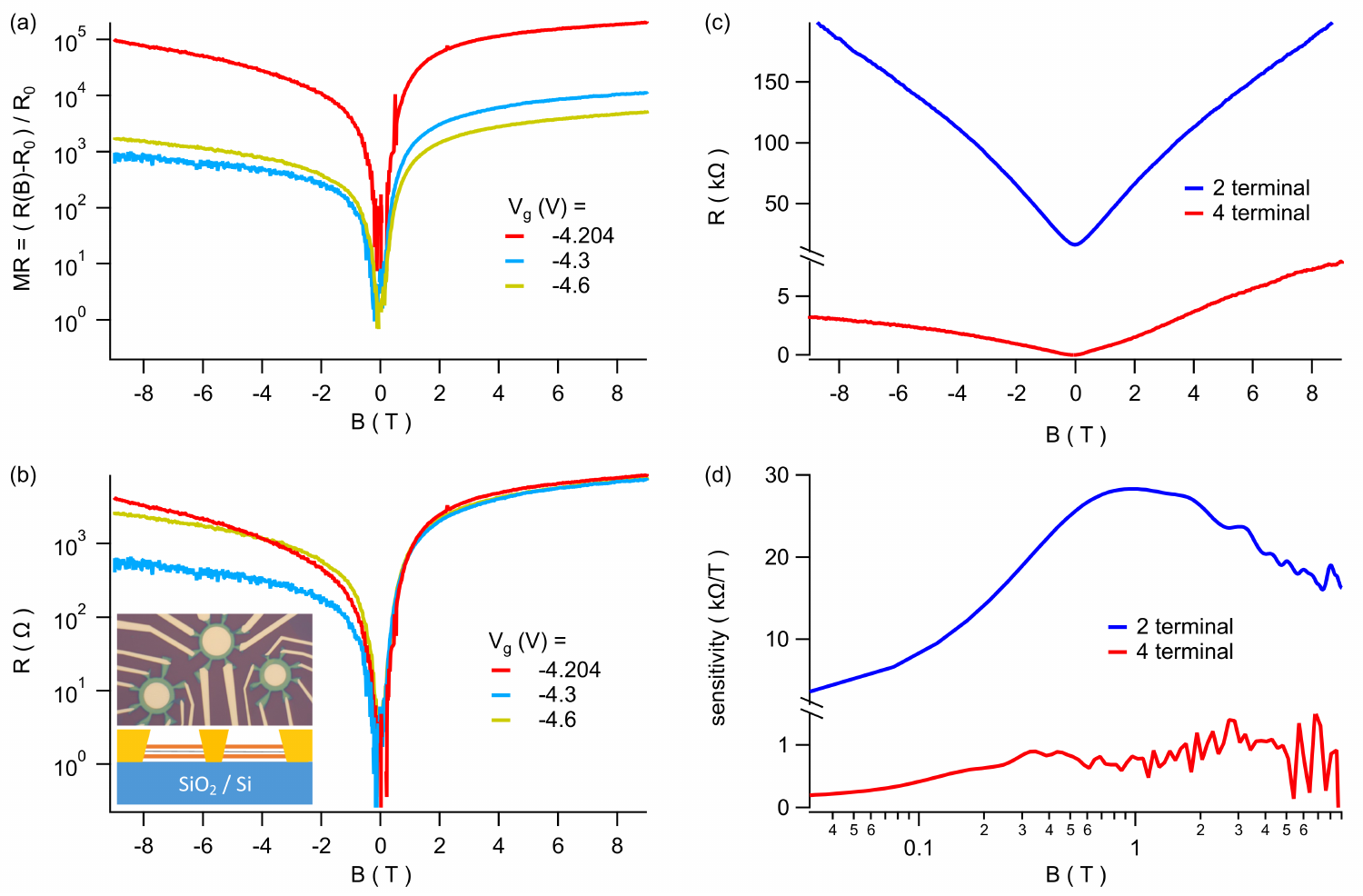}
\caption{(a) Magnetoresistance, $(R(B)-R_0 )/ R_0$, for the device with highest observed EMR effect at room temperature. The MR shows a strong dependence on back gate voltage. (b) The as-measured (un-normalized) resistance for the same gate voltages as in (a). The resistance itself shows little change with $V_g$. Inset shows on top: microscope image of a set of three devices fabricated from a single graphene/hBN stack; the yellow scale bar is 10 $\mu$m long. Contacts and central metallic shunt are made by edge contacts to etch-exposed graphene. Schematic shows side view of device geometry, with a central metal shunt flanked by metallic contacts to the edge. Metal is yellow, boron nitride is orange, graphene is thin black line. (c) Two resistance measurements for the same device, at $V_g{=}{-}4.2$ V. The red (blue) trace was acquired in a four- (two-) terminal configuration (red corresponds to same measurement in each figure). (d) The sensitivity, $dR/dB$, for the positive-field data in (c). Note the log-scale of the $B$-field axis. \label{fig1}}
\end{figure*}

The EMR devices here are made of flakes of monolayer graphene sandwiched between ${\sim}30$-nm-thick hBN flakes, assembled into stacks by a dry-transfer technique \cite{wang_one-dimensional_2013}. The device geometry is defined by reactive ion etching to create a disk with outer radius $r_b$ having a concentric circular hole of radius $r_a$. Films of Ti/Al, 4/80 nm thick, produce electrical contacts to the outer edge and a central metal shunt contacting graphene around the inner perimeter, as shown inset to Fig.\ \ref{fig1}(b). Electronic transport measurements were made by standard lockin techniques with a $200\ \mu$V bias at 13 Hz, at 300 K in a 9 T solenoid. Unless otherwise specified, all data is acquired in a four-terminal configuration. The graphene carrier density $n$ and  conductivity $\sigma$ are precisely controlled by  a gate voltage $V_g$ applied to the  Si substrate. The density, $n=6.7 \times 10^{10}\times \textrm{V}_g\ $V$^{-1}$cm$^{-2}$, is calibrated by  measurements in nearby Hall devices lacking the central shunt. Varying ratios of the metallic shunt to outer radius are used with  $r_a/r_b{=}0$ corresponding to a  device without the metal shunt. 

Figure\ \ref{fig1}a shows $MR$ from the device with the largest observed effect, for three closely spaced gate voltages which, though corresponding to very small changes in the carrier density of the device, nonetheless exhibit a remarkable variation in magnitude of the EMR. The high $MR$ ratio of ${\sim}10^5$ exceeds that of bulk 3D semiconductor devices (for a circular geometry; alternative geometries may yield much higher values \cite{pugsley_extraordinary_2013}). We also show in Fig.\ \ref{fig1}(b) the as-measured resistance, $R(B)$, for the same three traces as Fig.\ \ref{fig1}(a) showing that, at least for positive $B$-field, the $R(B)$ traces overlap almost identically. Thus, the variation of $MR$ with $V_g$ must be due to changes in the value of $R_0(V_g)$. For reference below, this device had an outer radius of $4.5\ \mu$m and a radius ratio of 0.85.

\begin{figure*}
\includegraphics[width=0.9\textwidth]{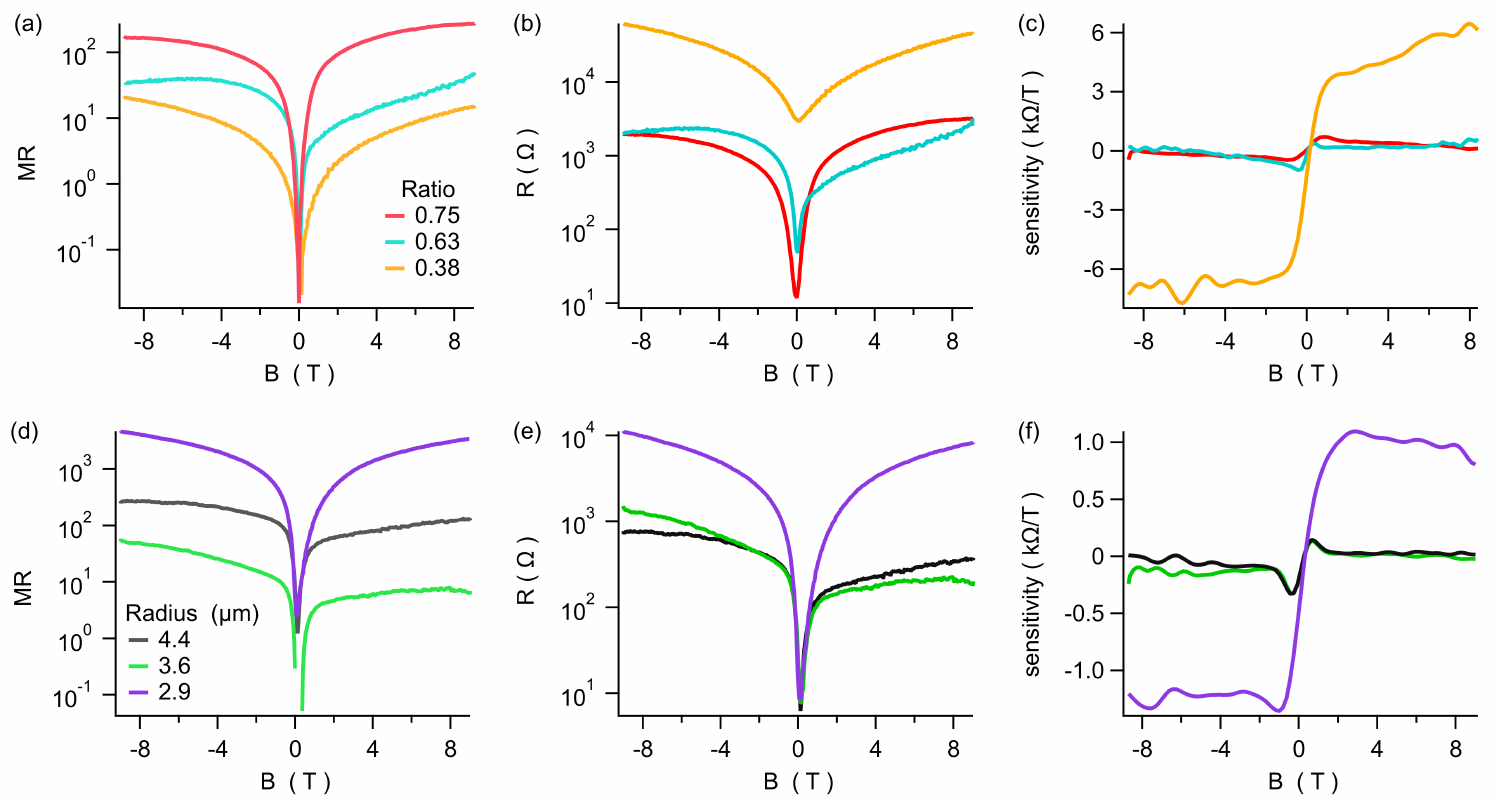}
\caption{The EMR properties of $MR$ ratio, measured resistance, and four-terminal sensitivity of devices with varying shunt-to-outer radius ratios but fixed outer radius of $2.9\ \mu$m  (a-c), and with varying outer radius at fixed ratio $r_a/r_b{=}0.75$ (d-f). Note the log scales of all plots except the sensitivity. Data in top and bottom set were acquired at different $V_g$, so traces with identical ratio and radius do not match. \label{fig2}}
\end{figure*}

\begin{figure*}
\includegraphics[width=0.9\textwidth]{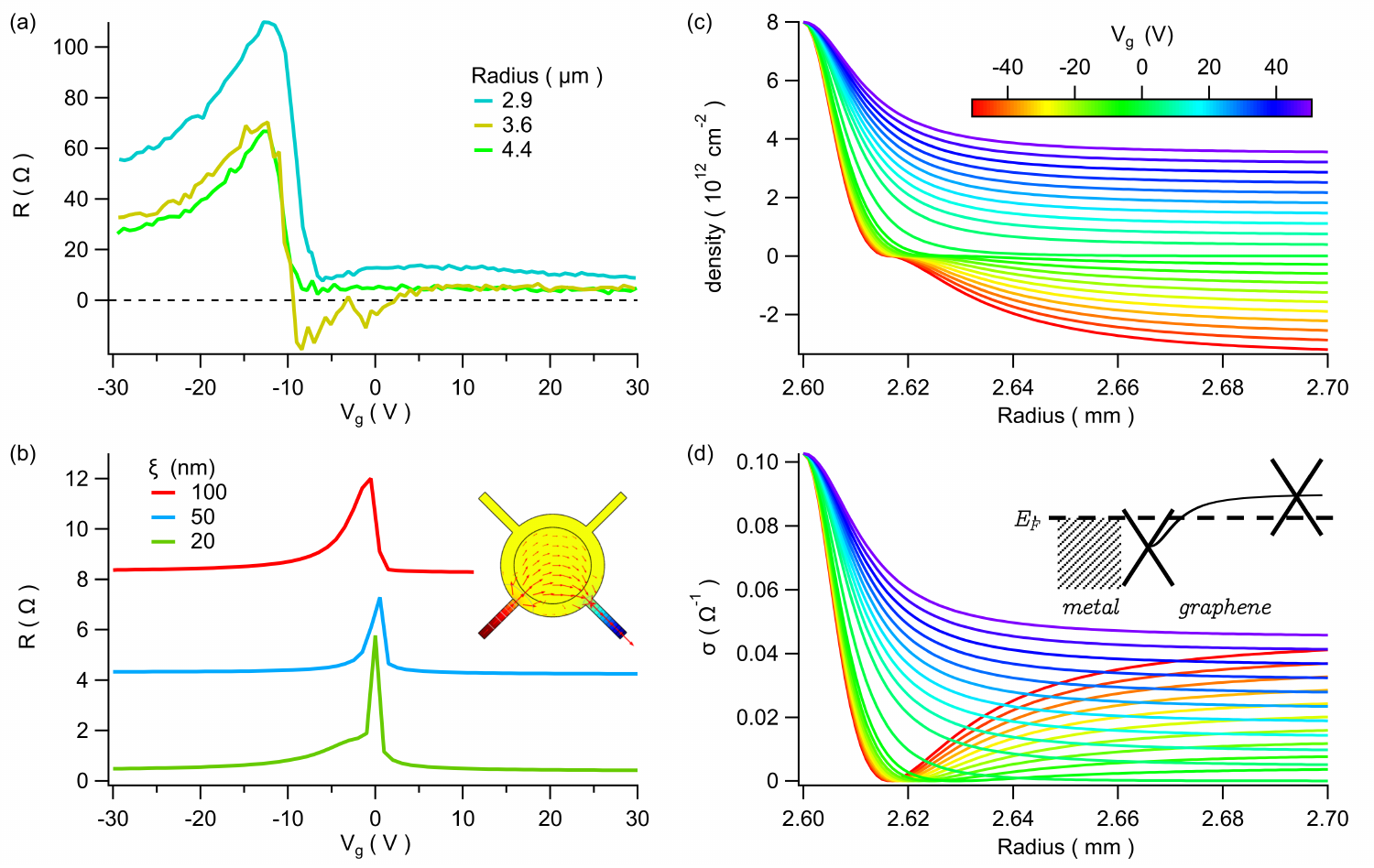}
\caption{(a) Asymmetry in the gate-voltage-dependent resistance at $B{=}0$ T and room temperature, in three devices of varying radii. (b)-(d) Results from a model of the conductivity in the neighborhood of the metal-graphene interface. (b) Model prediction for gate-voltage-dependent resistance at zero field; details in the main text. The traces are offset for clarity. Inset: Geometry of the EMR device used in finite element simulation. Both the graphene charge carrier density (c) and the conductivity (d) vary spatially near the metal-graphene interface, but in the bulk are determined primarily by the applied gate voltage. Inset to (d) shows a schematic of the graphene Dirac cone position relative to the Fermi energy, both near to and far from a metallic contact, for an instance when the bulk graphene has been gated to be $p$-type. \label{fig3}}
\end{figure*}

Typically, circular EMR devices are measured using four contacts spaced at 90$^{\circ}$ relative to each other, and where possible this standard was followed here (e.g.\ in Fig.\ \ref{fig1}). In several devices, design constraints or lost contacts led to the use of  contacts clustered within less than half the circumference of each device  (as for data in Fig.\ \ref{fig2}). The metallic shunt is not always concentric with the outer device radius. Such non-idealities can lead to asymmetry in the EMR \cite{solin_extraordinary_2001}, and may  be responsible for the observed asymmetry between positive and negative magnetic fields. However, this  does not alter our findings:\ the results of Fig.\ \ref{fig1} are obtained in the standard geometry, while those of Figs.\ \ref{fig2} and \ref{fig3} are comparisons between samples with similar contact configurations. 

While the $MR$ ratio is a standard figure-of-merit, the sensitivity $dR/dB$ may be more useful as it captures the field-dependent response of the device, and is not sensitive to $R_0$. This is helpful for devices that 1) have a nonlinear response to $B$, and 2) have a gate-dependent $R_0$. Thus in addition to the $MR$, we also plot in Fig.\ \ref{fig1}(c) and d the measured $R$ and corresponding $dR/dB$ (determined by numerical differentiation, which accounts for the high frequency noise in Fig.\ \ref{fig1}(d)\ ) in a single device, the same as for Fig.\ \ref{fig1}(a) and (b), at $V_g=-4.2$ V. Results are shown for both a 2- or 4-terminal configuration. Two features stand out: first, the 2-terminal sensitivity is nearly twenty times larger than previously reported for graphene \cite{lu_graphene_2011} and sixty times larger than the highest reported value  in semiconductor structures \cite{sun_strong_2012}. Second, the 2-terminal sensitivity greatly exceeds the 4-terminal values in the same device at the same $V_g$. The reason has been previously discussed \cite{sun_design_2012} and a smaller instance was observed in Ref.\ \cite{lu_graphene_2011}. The idea is straightforward: the farther the voltage probes are from the current contacts, the smaller the measured potential drop since the shunt enforces an equipotential at its edge. Therefore when the voltage probes approach and merge with the current contacts, the potential drop in the resulting two-terminal geometry reaches a maximum.  These sensitivities exceed those of Hall sensors based on encapsulated graphene devices \cite{dauber_ultra-sensitive_2015}. We note that encapsulated graphene devices are a technology that is still improving rapidly, with recent advances in limiting the role of scattering by using nearby graphite gates \cite{zibrov_tunable_2017}, and implementing 2D contacts by XeF$_2$ etching of the hBN layers \cite{son_atomically_2018} which may greatly reduce the contact resistance and improve contact reproducibility. Neither approach is employed in the present study, suggesting that further improvements in device performance are achievable. 

We use this result to estimate the achievable field resolution, $\delta B = \delta R/(dR/dB)$. The resistance resolution $\delta R$ is estimated to be the standard deviation of the  measured resistance due to noise. For the 2- and 4-terminal data, $\delta R \approx 120\ \Omega$ and $20\ \Omega$, respectively, leading to estimated peak resolutions of $\delta B_{2pt} = 4$ mT and $\delta B_{4pt} = 30$ mT with a one second averaging time.  These values can be improved by e.g.\ longer meaurement times or impedance-matched preamplifiers.

In Figure\ \ref{fig2} we compare the $MR$ ratio, measured resistance, and four-terminal sensitivity for two sets of devices. Panels (a-c) show results for a fixed outer radius of $2.9\ \mu$m and a varying shunt-to-outer radius ratio $r_a/r_b$, while in Fig.\ \ref{fig2}(d)-(f) the ratio is held at 0.75 and the outer radius is varied. For all traces the gate voltage was held near charge neutrality of the graphene. The EMR is largest for the highest ratio, similar to prior work \cite{solin_extraordinary_2001}. However, the largest device resistance and sensitivity are found for the \emph{smallest} ratio. This highlights the role of $R_0$ in determining the magnitude of $MR$, while the sensitivity responds to the steepest change of $R(B)$ which is independent of $R_0$. The data for varying the overall radius is less conclusive: while the device with the smallest radius gives the largest response across the board, the other traces do not reveal a clear size dependence.

In principle an advantage of graphene-based EMR devices is the inherent tunability of graphene via $V_g$. In graphene devices without a shunt, the resistance reaches a maximum at charge neutrality and falls off for increasing electron or hole charge density \cite{novoselov_electric_2004}. In shunted graphene-on-oxide devices, the maximum survives but is rather broad \cite{pisana_graphene_2010,pisana_tunable_2010,lu_graphene_2011,friedman_extraordinary_2011}. In contrast, here the resistance maximum survives but invariably shows a strong asymmetry near charge neutrality; equally important, the resistance can become \emph{negative}. An example is shown in Fig.\ \ref{fig3}(a), showing zero-field transport in three devices having the same ratio but different outer radii. In encapsulated graphene devices, the carrier mean free path  can be $\sim \mu$m even at room temperature \cite{wang_one-dimensional_2013}, approximately the spacing between the shunt and outer radius of our devices. Thus transport becomes quasi-ballistic and ``negative'' resistances can arise due to non-diffusive transport \cite{wang_one-dimensional_2013,solin_nonmagnetic_2002}. In such a device $R_0$ can be tuned to zero, yielding an infinite $MR$; this is a key reason for using sensitivity as a figure of merit.

The origin of the asymmetry in the zero-field resistance vs $V_g$ is likely due to the band re-alignment that occurs near a metal-graphene interface. This issue has been extensively investigated in the context of electrical contacts to graphene\ \cite{giovannetti_doping_2008,khomyakov_first-principles_2009,mueller_role_2009,matsuda_contact_2010,xia_origins_2011,allain_electrical_2015}. Briefly, the large density of states in the metal pins the Fermi level in graphene so it is highly doped (with most metals yielding $n$-type doping) near the contact. But in the bulk of graphene, $n$ is still controlled by $V_g$. Since the polarity at the metal-graphene interface is fixed, a $pn$-junction arises near the contacts depending on the charge state in the graphene bulk, effectively increasing the shunt contact resistance. Thus an asymmetry in the device behavior is expected as a function of $V_g$.

To verify this picture we model transport through an EMR device assuming an edge-contacted geometry for the metal-graphene interface. Here one expects better overlap of the metal and carbon orbitals than for surface-contacted graphene, where the metal atoms encounter graphene $\pi$ orbitals that will not form covalent bonds \cite{matsuda_contact_2010}. In models of surface-contacted graphene, this weak coupling allows the graphene Fermi level to move in response to $V_g$, even for graphene directly under the metal \cite{xia_origins_2011}. In the present case, we assume the graphene density is pinned by the metal at the interface, no matter the $V_g$ in the bulk. Thus the potential relaxes over a characteristic distance $\xi$ from the interface  back to the bulk value set by $V_g$. Following Xia \etal\ \cite{xia_origins_2011}, the potential relaxation is
\begin{equation*}
E_F(r)- E_{F,b} =\left(\Delta W - E_{F,b} \right) \left(1+\left( 3 (r-r_a)/\xi \right)^2\right)^{-1}\ ,
\end{equation*}
\noindent where $r$ is the radial position, $E_F(r)$ the position dependence of the Fermi energy relative to charge neutrality in graphene, $E_{F,b}$ is the $V_g$-controlled Fermi energy in the bulk, and $\Delta W$ is the work function difference of the metal and graphene. We use $\Delta W{=}0.33$ eV  \cite{khomyakov_first-principles_2009}, and calculate the carrier density profile $n(r){=}(E_F(r)/ \hbar v_F)^2/\pi$ with the band velocity $v_F{=}10^6$ m/s. We then use finite element simulation to determine the total resistance of an EMR device with a metallic shunt and a gate-voltage-dependent graphene conductivity that varies spatially with the density as $\sigma(r)=|n(r)| e \mu + \sigma_{min}$.

In Fig.\ \ref{fig3}(b) we show the results of this simple model, plotting the predicted zero-field resistance vs applied gate voltage for three values of the width $\xi$. A clear similarity to the lineshape of the data in Fig.\ \ref{fig3}(a) is achieved, with best results for $\xi{=}100$ nm: the asymmetry appears, over a gate voltage range not much less than we observe, and the variation in resistance is, if not close, at least of similar magnitude. The most significant departure is in the $V_g$ value at the resistance peak. However, the peak location is sensitive to charge doping by extrinsic sources which we do not attempt to capture in this model. Figure \ref{fig3}(c) and (d) show the density and conductivity profiles for a range of $V_g$, in a model geometry with $r_a{=}2.6 \mu$m. As expected, for $V_g{<}0$ when the graphene bulk is $p$-type, the conductivity reaches a minimum a short distance away from the graphene-metal interface due to the $pn$-junction. Thus the overall device resistance reflects the fact that the low-impedance path through the shunt is effectively shielded by the $pn$-junction when the bulk of graphene is doped opposite to that induced by the metal interface.

In conclusion, we have investigated the EMR effect in encapsulated graphene devices. We find the magnetoresistance can be enhanced by over four orders of magnitude from its zero field value, with an $MR$ ratio that can surpass the highest reported for a circular device geometry. We also find very large values of the sensitivity, $dR/dB$, reaching nearly $30$ k$\Omega$/T. An asymmetry in the gate voltage response of the zero-field resistance is traced to the presence of $pn$-junctions near the graphene-metal interface, and a model of edge-contacted graphene is in reasonable agreement with these observations. Encapsulated graphene is thus a promising platform for high-sensitivity measurements of magnetic fields using the EMR effect.

\begin{acknowledgments}
We thank S.\ Solin for inspiration and a close reading of the manuscript, and gratefully acknowledge support from the Institute of Materials Science and Engineering at Washington University. EAH acknowledges partial support under NSF DMR-1810305.
\end{acknowledgments}


%

\end{document}